\documentclass[letterpaper,english,aps,letterpaper,prl,preprintnumbers,twocolumn]{revtex4}
\usepackage[latin9]{inputenc}
\usepackage{pifont}

\makeatletter
\usepackage{graphicx}
\usepackage{dcolumn}
\usepackage{bm}
\usepackage{hyperref}
\usepackage[usenames]{color}
\makeatletter \def\@dotsep{5} \makeatother
\usepackage{mathptmx, courier, pifont}
\usepackage[T1]{fontenc}
\usepackage{textcomp}
\usepackage{bm}
\usepackage{amsmath}
\usepackage{amssymb}

\newcommand{\singlefig}[6]{%
\begin{figure}\vspace{#3}%
\includegraphics*[scale=#5]{#2}%
\caption{\label{fig:#1} #6}%
\vspace{#4}%
\end{figure}}

\renewcommand{\vec}{\mathbf}

\AtBeginDocument{

}

\usepackage{babel}
\makeatother

\begin{document}

\title{Valence bond glass --- A unified theory of electronic
disorder and pseudogap phenomena \\ in high T$_c$ cuprate
superconductors}

\author{Liang Ren Niestemski and Ziqiang Wang}

\affiliation{Department of Physics, Boston College, Chestnut Hill,
Massachusetts 02467}

\begin{abstract}
We show that the low-energy fluctuations of the valence bond in
underdoped high-T$_c$ cuprates, originating from quantum
fluctuations of the superexchange interaction, are pinned by the
electronic disorder due to off-stoichiometric dopants, leading to a
valence bond glass (VBG) pseudogap phase. The antinodal Fermi
surface sections are gapped out, giving rise to a normal state Fermi
arc whose length shrinks with underdoping.
%below a temperature $T^*$ determined by thermal filling.
Below $T_c$, the superexchange interaction induces a $d$-wave
superconducting gap that coexists with the VBG pseudogap. The
evolution of the local and momentum-space spectroscopy with doping
and temperature captures the salient properties of the pseudogap
phenomena and the electronic disorder.
% revealed by recent ARPES and
% STM experiments.
% in underdoped and chemically substituted high-T$_c$ superconductors.
The unified theory elucidates the important interplay between strong
correlation and the intrinsic electronic disorder in doped
transition metal oxides.
\end{abstract}


\maketitle

The least understood electronic state in the high-T$_c$ cuprates
lies in the pseudogap phase that straddles the AF Mott insulator and
the $d$-wave superconductor. The origin of the pseudogap phase is a
subject of intensive debate \cite{leenagaosawen}.
%in connection to the emergent quantum electronic matter in doped Mott insulators
This unexpected phase is defined by a normal state {\em pseudogap}
in the single-particle excitation spectrum below a characteristic
temperature $T^*$ \cite{loeser,hding}. One of the most intriguing
properties is the momentum space anisotropy: the Fermi surface (FS)
near the $d$-wave antinodes is gapped out, leaving a Fermi arc of
gapless excitations near the nodes. This nodal-antinodal dichotomy
is further enlightened by recent ARPES \cite{tanaka06,lee07,kondo07}
and Raman \cite{letacon} experiments that find two distinct spectral
gaps in terms of the temperature and doping dependence: a large
pseudogap near the antinodes that tracks $T^*$ and a smaller
superconducting (SC) gap along the arc that tracks $T_c$. Recent STM
measurements observe a SC gap coexisting spatially with a large
pseudogap below $T_c$ \cite{boyer,vidya}.

The Fermi arc raises the possibility for an unconventional
electronic state. In the absence of disorder, the FS is the
trajectory of the poles in the single-particle Green's function
(discontinuity in the momentum distribution) and involves continuous
contours in momentum space. Thus the Fermi arc must evolve into
either a Fermi point or a FS pocket at low temperatures. This
difference highlights two different (one-gap and two-gap) proposals
for the origin of the pseudogap. In the one-gap scenario, the
pseudogap is a $d$-wave pairing gap without SC phase coherence that
evolves into a single SC gap below $T_c$
\cite{leenagaosawen,normanreview,arpes-scaling}. The two-gap
scenario attributes the pseudogap as due to competing order from an
energetically favorable state unrelated to pairing. An important
implication of a generic two-gap scenario is the {\em coexistence}
of a $d$-wave SC gap over the Fermi arc and a low-temperature
pseudogap in the antinodal region below $T_c$ \cite{civelli}. The
observation of two distinct gaps by ARPES, Raman, and STM
experiments provides strong support for the two-gap scenario.

A consistent theory for the pseudogap has been challenging. Most of
the proposals for a competing state invoke various density waves and
flux/orbital current order \cite{leenagaosawen,normanreview} that
break translation symmetry by one lattice constant and result in
superstructures. The FS is truncated in the antinodal region by
band-folding with respect to a unique ordering wavevector $q$,
leading to FS pockets and particle-hole asymmetric pseudogap density
of states \cite{cli06}. However, despite many years of effort and
the much improved measurement resolution, no signatures of disguised
FS pockets or folded bands have been detected in zero external
fields.
%Moreover, the pseudogap produced by these density wave orders is
%particle-hole asymmetric \cite{cli06}.
%, which is inconsistent with the V-shaped averaged density of states
%(DOS) at low energies observed in tunneling experiments.

A related phenomenon in the cuprates is the electronic disorder.
There has been mounting STM evidence for nanoscale DOS gap disorder
\cite{shpan,howald,lang,mcelroy05,hanaguri} and, under a variety of
conditions where superconductivity is weakened, short-range ordered
checkerboard DOS modulations
\cite{hanaguri,howald,vershini04,mcelroy-prl05} which are
manifestations of a bond-centered electronic glass \cite{kohsaka07}.
A natural cause for the electronic disorder is the out of plane
ionic dopants, interstitial in Bi$_2$Sr$_2$CaCu$_2$O$_{8+x}$,
substitutional in La$_{2-x}$Sr$_x$CuO$_4$ and
Ca$_{2-x}$Na$_x$CuO$_2$Cl$_2$, and in combination with chemical
substitutions in Bi$_2$Ln$_{2-z}$Bi$_z$CuO$_{6+x}$ (Ln-Bi2201). In
addition to inducing structural distortions, the screening of the
dopant electrostatic potential is highly nonlinear in doped Mott
insulators due to strong Coulomb repulsion, leading to inhomogeneous
electronic states with spatial variations in the local doping
concentration \cite{wang02,zhoudingwang}.

We present here a unified theory for the pseudogap phenomena and the
electronic disorder. The key point is the interplay between strong
correlation induced valence bond fluctuations and the dopant induced
disorder. In the parent compounds, the superexchange interaction
between the Cu spins, described by the nearest neighbor Heisenberg
model $H_J= J\sum_{\langle i,j\rangle} {\bf S}_i\cdot{\bf S}_j$,
causes AF order in the ground state. Due to strong quantum
fluctuations of the spin-1/2 moment, the nonmagnetic {\em valence
bond states} are close in energy to the AF state. Writing ${\bf
S}_i\cdot{\bf S}_j={1\over2}c_{i\sigma}^\dagger
c_{i\sigma^\prime}c_{j\sigma^\prime}^\dagger c_{j\sigma}+{\rm
const.}$, the valence bond can be formed via spin-singlet pairing
$\Delta_{ij}=\langle
c_{i\uparrow}c_{j\downarrow}-c_{i\downarrow}c_{j\uparrow}\rangle$ as
in the resonance valence bond (RVB) theory \cite{andersonRVB}, and
the paramagnetic $\chi_{ij}=\sum_\sigma\langle c_{i\sigma}^\dagger
c_{j\sigma}\rangle$ as envisioned by Pauling in chemical bonding.
Since charge fluctuations are completely suppressed, the two
descriptions are equivalent and the valence bond states are highly
degenerate owing to the SU(2) symmetry \cite{su2}. Besides valence
bond liquid states, there are also symmetry breaking valence bond
crystal states that are gapped but competitive in energy
\cite{leenagaosawen,kivelson-rhoksar}. Doping the Mott insulator
breaks the SU(2) symmetry and makes the valence bond in the
particle-particle and particle-hole channels different. The basic
question is which fluctuating valence bond state is selected when a
sufficient amount of holes destroys the AF long-range order. In the
short-range RVB theory, the spin-singlet pairs are mobilized by the
doped holes and condense into a $d$-wave SC state \cite{gangoffive}.
There is, however, a natural competing order driven by the same
superexchange interaction, i.e. the paramagnetic valence bond
$\chi_{ij}$.

We show that the $d$-wave component of the {\em real part} of
$\chi_{ij}$ represents the most important low-energy fluctuations
and that dopant induced disorder pins such $d$-wave charge density
wave fluctuations to an electronic valence bond glass (VBG) phase
exhibiting the observed pseudogap phenomena. This is similar in
spirit to disorder induced glassy phases or the nematic liquid
crystal of stripes \cite{kivelson-rmp,vojta}. A ubiquitous feature
of the VBG is the emergence of a pseudogap near the antinodes where
the valence bond fluctuations are large due to the flatness of the
band, giving rise to a genuine normal state Fermi arc. Based on
microscopic calculations of the extended t-J model using spatially
unrestricted Gutzwiller approximation, we show that the VBG captures
the salient properties of the observed pseudogap phenomena and
electronic disorder, especially in underdoped and chemically
substituted bilayer and single-layer Bi-based cuprates where the
pseudogap and the SC gap are well separated
\cite{tanaka06,lee07,kondo07,boyer,vidya}.

The Hamiltonian of the extended t-J model is given by
\begin{equation}
H=-\sum_{i\neq j}t_{ij}P_ic_{i\sigma}^\dagger c_{j\sigma}P_j
+J\sum_{\langle i,j\rangle}({\bf S_i}\cdot{\bf S_j} -{1\over4}\hat
n_i\hat n_j)
%+\sum_{i\neq j}\hat n_i V_{ij}^c \hat n_j
+\sum_{i}V_{i}\hat n_{i}. \label{h}
\end{equation}
The electrons hop between near neighbors via $t_{ij}$. Repeated spin
indices are summed, $\hat n_i=c_{i\sigma}^\dagger c_{i\sigma}$,
${\bf
S}_i={1\over2}c_{i\sigma}^\dagger{\vec\tau}_{\sigma\sigma^\prime}
c_{i\sigma^\prime}$.
%The local doping is $x_i=1-n_i$ with
%$n_i=\langle c_{i\sigma}^\dagger c_{i\sigma} \rangle$.
% and the average doping $x=(1/N_s)\sum_i x_i$ on a square
% lattice of $N_s$ sites with lattice constant set to unity.
The projection operator $P_i$ removes double occupation on site-$i$.
The last term in Eq.~(\ref{h}) is the electrostatic potential,
$V(i)=\sum_{j\neq i}{V_c\over\vert r_i-r_j\vert}(n_j-{\bar
n})+\sum_{\ell=1}^{N_d}{V_d\over\sqrt{\vert
r_i-r_{\ell}\vert^2+d_s^2}}$, where the long-range Coulomb
interaction of strength $V_c$ between the in-plane electrons
provides the important screening of the ionic potential of strength
$V_d$ from $N_d$ number of off-plane dopants at a set back distance
$d_s$ \cite{wang02,zhoudingwang}. We use $J=120$meV and up to fifth
nearest neighbor hoppings $t=(360,-120,29,24,-24)$meV relevant for
the band structure \cite{normanding} and set $V_c=V_d=0.5$eV and
$d_s=1$ in units of the lattice constant \cite{wang02}. To account
for their Coulomb repulsion, the ionized dopant configurations are
generated randomly with a hard-core of one to three lattice
spacings. \singlefig
     {label} % Label
           {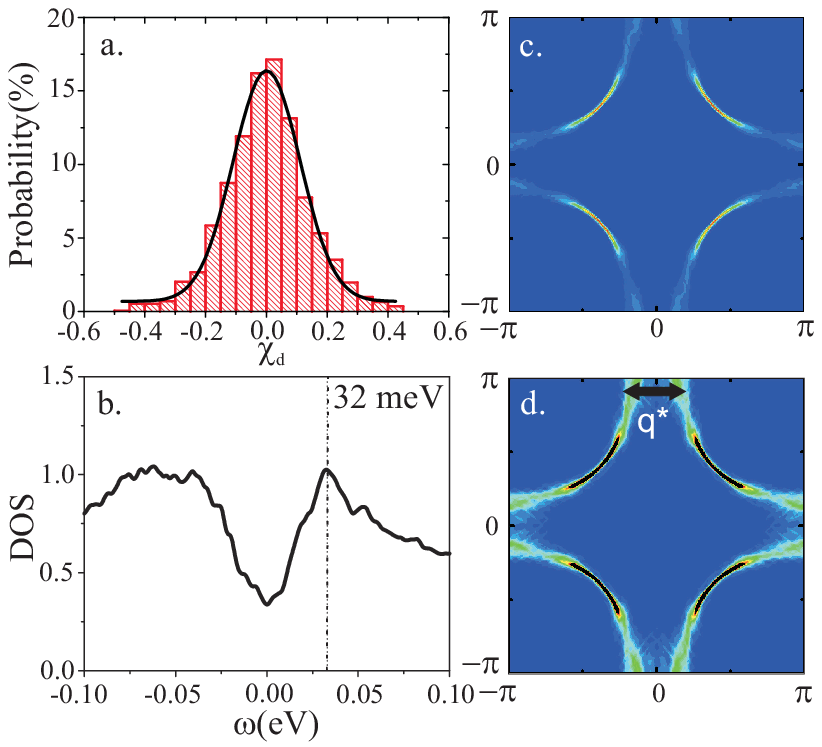} % Filename (with path if needed)
           {0pt} % Extra space above (with units)
           {0.0pt} % Extra space below (with units)
           {1.0} % Scale factor (1.0 is full size)
           {Normal state pseudogap and Fermi arc at $x=0.125$.
           (a) Distribution of $d$-wave valence bond showing glassy order.
           (b) Averaged DOS showing VBG pseudogap.
           (c) and (d) Fermi arc.} % Figure caption

To account for both strong correlation and disorder, we use the
spatially unrestricted Gutzwiller projected wave function
%\begin{equation}
$\vert\Psi\rangle=\prod_i y_{i}^{\hat n_{i}}(1-{\hat n_{i\uparrow}}
{\hat n_{i\downarrow}})\vert\Psi_0\rangle$,
% \label{wf}
%\end{equation}
where $\Psi_0$ is a Slater determinant state and $y_{i}$ is a {\em
local} fugacity that keeps the density unchanged before and after
projection. The projection is implemented using the Gutzwiller
approximation \cite{gutzwiller,zhang}. The t-J Hamiltonian is
replaced by one in the unprojected space with renormalized hopping
and exchange that capture the basic Mott physics: $t_{ij}\to
g_{ij}^t t_{ij}$ and $J\to g_{ij}^J J$. In an inhomogeneous state,
%\begin{equation}
$g_{ij}^t=
%{\langle\Psi\vert c_{i\sigma}^\dagger
%c_{j\sigma}\vert\Psi\rangle\over\langle\Psi_0\vert
%c_{i\sigma}^\dagger c_{j\sigma}\vert\Psi_0\rangle}=
\sqrt{4x_i x_j/(1+x_i)(1+x_j)}, \ g_{ij}^J=
%{\langle\Psi\vert {\bf
%S}_i\cdot {\bf S}_j\vert\Psi\rangle\over\langle\Psi_0\vert
%c_{i\sigma}^\dagger c_{j\sigma}\vert\Psi_0\rangle}=
{4/(1+x_i)(1+x_j)}
%\label{gfactor}
%\end{equation}
$ depend on the {\em local doping} $x_i$ \cite{cli06,zhoudingwang}.
Decoupling the exchange interaction in terms of the valence bond
$\chi_{ij}$ and singlet pairing $\Delta_{ij}$, we obtain a
Gutzwiller renormalized Hamiltonian,
\begin{eqnarray}
H_{\rm GA}= &-&\sum_{i\neq j}g_{ij}^t t_{ij}c_{i\sigma}^\dagger
c_{j\sigma} +\sum_i(V_i+\lambda_i)c_{i\sigma}^\dagger
c_{i\sigma}-\sum_i\lambda_i n_i
\nonumber \\
&-&{1\over4}J\sum_{\langle i,j\rangle}g_{ij}^\chi
\left(\chi_{ij}^*c_{i\sigma}^\dagger c_{j\sigma}+{\rm
h.c.}-\vert\chi_{ij}\vert^2\right) \nonumber \\
&-&{1\over4}J\sum_{\langle
i,j\rangle}g_{ij}^\Delta\left(\Delta_{ij}^*\epsilon_{\sigma\sigma^\prime}
c_{i\sigma} c_{j\sigma^\prime}+{\rm
h.c.}-\vert\Delta_{ij}\vert^2\right), \label{hga}
\end{eqnarray}
where $g_{ij}^\chi=g_{ij}^\Delta=g_{ij}^J$ and $\lambda_i$
originates from the local fugacity. We minimize the ground state
energy of Eq.~(\ref{hga}) through self-consistently determined
$\{x_i,\lambda_i,\chi_{ij},\Delta_{ij}\}$ on $24\times24$ systems
for different dopant configurations.

{\em Normal state VBG pseudogap phase}. In the normal state above
$T_c$ or the zero temperature phase with $\Delta_{ij}=0$, we find
that the valence bond $\chi_{ij}$ is {\em real} and fluctuates due
to the disorder potential in the doping range studied. The dominant
fluctuation is in the $d$-wave channel, $\chi_d(i)={1\over4}\sum_j
d_{ij}\chi_{ij}$, with the form factor $d_{ij}=\pm1$ for the four
bonds emanating from site $i$. We first focus on the normal state by
setting $\Delta_{ij}=0$. The histogram of $\chi_d$ at average doping
$x=0.125$ is shown in Fig.~1a. It follows a Gaussian distribution
with zero mean. The root-mean-squared fluctuation represents a
nonzero ``glassy'' order parameter
$\delta_\chi=\sqrt{\sum_i\chi_d^2(i)}/N_s =0.28$ for the VBG phase.
We note that glassy dynamics of valence bond was studied in the
weak-coupling metallic phase of the Hubbard-Heisenberg model at
half-filling \cite{tarzia}. The most succinct feature of the VBG is
the emergence of the pseudogap and the Fermi arc. The averaged DOS
in Fig.~1b shows a remarkable V-shaped pseudogap, $\Delta_{pg}\sim
32$meV, approximately symmetrically distributed around the Fermi
level due to the $d$-wave nature of the VBG. The calculated spectral
intensity at the Fermi energy in Fig.~1c reveals a FS truncated near
the antinodes by the pseudogap and a Fermi arc around the nodes. The
Fermi arc tracks the underlying FS and remains prominent upon
lowering the intensity scale in Fig.~1d without signs of band
folding.
 \singlefig
     {label} % Label
           {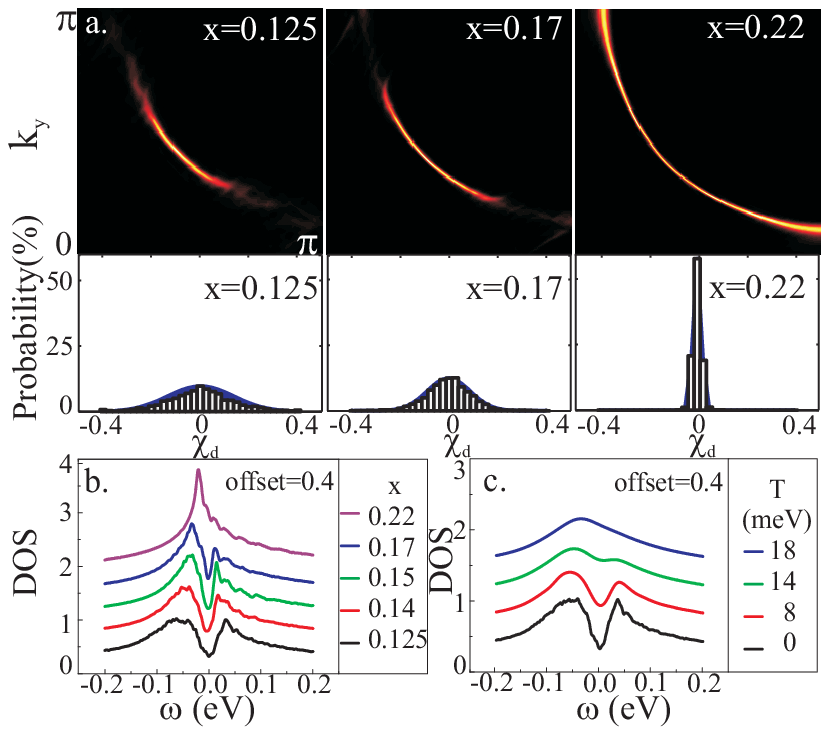} % Filename (with path if needed)
           {0pt} % Extra space above (with units)
           {0.0pt} % Extra space below (with units)
           {1.0} % Scale factor (1.0 is full size)
           {Doping dependence of the Fermi arc and
           valence bond glassy order (a) and averaged
           DOS and pseudogap (b). (c) Temperature evolution of VBG
           pseudogap at $x=0.125$.} % Figure caption

The doping and temperature dependence of the VBG pseudogap is shown
in Fig.~2. With increasing doping, the distribution of the valence
bond in Fig.~2a sharpens around zero, the VBG order parameter
$\delta_\chi$ reduces, and the Fermi arc length increases in
agreement with ARPES experiments. Note that while the density of
off-plane dopants increases with doping, $\delta_\chi$ decreases due
to improved screening by more mobile carriers that leads to weaker
fluctuations of the potential $V_i$. The doping evolution of the
averaged DOS is shown in Fig.~2b. The pseudogap becomes smaller and
shallower with increasing doping and becomes undiscernible beyond
$x=0.22$ on the shoulder of the van Hove peak. The extracted
pseudogap has a doping dependence that follows
$\Delta_{pg}\simeq(J/8)\sqrt{\sum_{i}(\sum_jg_{ij}^\chi\chi_{ij}d_{ij})^2/N_s}$,
as shown in Fig.~3c. The temperature evolution of the pseudogap
obtained by minimizing the free energy of Eq.~(\ref{hga}) is shown
in Fig.~2c at $x=0.125$. The pseudogap onset temperature
$T^*\simeq16$meV is clearly seen to be determined by the thermal
filling of the VBG pseudogap.

The momentum-dependence of the symmetrized spectral function
$A(k,\omega)$ is plotted in Fig.~3a at Fermi level for $x=0.14$. It
shows gapless quasiparticle excitations along the arc and the
opening of the pseudogap at the arc tip that increases toward the
antinode. The angular dependence of the pseudogap extracted this way
is shown in Fig.~3b. Away from the zero-gap Fermi arc regime, the
pseudogap follows the $d$-wave form (dashed line) consistent with
the $d$-wave nature of the VBG. The line-shape in Fig.~3a shows that
the pseudogap near the antinodes is a {\em soft gap}. This important
prediction of the VBG is consistent with the original ARPES
experiments \cite{loeser,hding} and differs from the naive picture
of a hard gap that depletes all states near the antinode below the
pseudogap energy scale. These in-gap states contribute to the
spectral weight at low energy and have important consequences in the
SC state.

The pseudogap in the VBG theory is not sensitive to the precise
topology of the FS and does not require the antinodal FS sections to
be parallel which can induce long-range incommensurate density wave
order \cite{cli06}. We obtain similar results using a different set
of hopping parameters $t=(480,-160,50,50,-50)$meV used in
Ref.\cite{zhoudingwang} that does not favor the antinodal nesting
condition.
% These will be used later for the SC state.
A band dispersion showing the observed flatness near the antinode
suffices. The propensity toward the VBG is due to an enhanced static
susceptibility broadly peaked around the $q^*$ shown in Fig.~1 that
enables electronic disorder to pin the VBG with a distribution of
$\chi_d(q)$ peaked around $q^*$. In Fig.~3d, we show the Fourier
power spectrum $\vert\chi_d(q)\vert$ plotted along the
$q_x$-direction. Both the peak and the incommensurate $q^*$ decrease
with increasing doping. At $x=0.14$, $q^*\simeq(\pm0.4\pi,0)$,
indicative of glassy $\sim 5\times5$ checkerboard patterns for the
valence bond.
 \singlefig
     {label} % Label
           {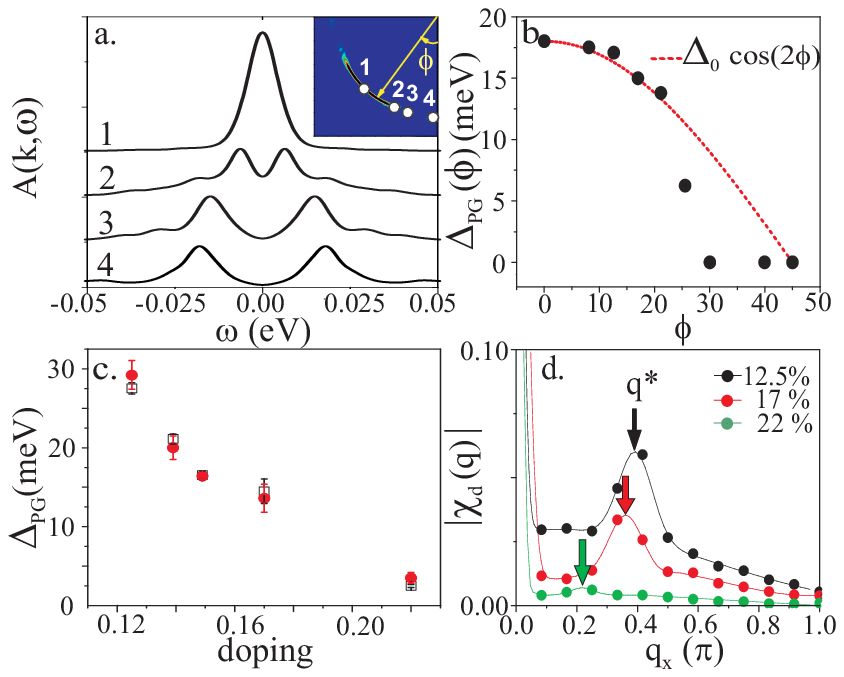} % Filename (with path if needed)
           {0pt} % Extra space above (with units)
           {0.0pt} % Extra space below (with units)
           {1.0} % Scale factor (1.0 is full size)
           {Angule dependence of
           symmetrized spectral function (a) and pseudogap (b).
           (c) Doping dependence of pseudogap from DOS (circles) and
           $d$-wave valence bond order parameter (squares). (d)
           $\vert\chi_d(q)\vert$ along $q_x$-direction with
           peaks at $q^*$ marked in Fig.~1d. } % Figure caption

{\em Superconducting phase}. The VBG can coexist with an
inhomogeneous $d$-wave superconductor; both due to the superexchange
interaction. Although the Gutzwiller factors
$g_{ij}^\chi=g_{ij}^\Delta$, charge fluctuations at finite doping,
and in particular, the pair-breaking induced by inter-site Coulomb
repulsion will weaken the singlet pairing channel \cite{zhouwang}.
Moreover, the electron-phonon interactions have been shown to
promote the $d$-wave charge density wave \cite{newns,dhlee}. To
incorporate these effects into the renormalized mean field theory in
Eq.~(\ref{hga}), we use $g_{ij}^{\chi}=g_{ij}^J$ and
$g_{ij}^\Delta=pg_{ij}^J$ with $p=0.475$, which separates the two
energy gap scales in the underdoped regime. The averaged DOS at
$T=0$ is shown in Fig.~4a for $x=0.14$. It displays two gaps: a
smaller SC gap $\Delta_{sc}\simeq10$meV and a large pseudogap
$\Delta_{pg}\simeq22$meV inherited from the normal state, in
agreement with STM experiments on La-Bi2201 \cite{boyer,vidya}. The
temperature evolution in Fig.~4b shows that the SC gap and coherence
peaks disappears above $T_c\sim4$meV as the system enters the
pseudogap phase. Interestingly, as $T$ is increased toward $T_c$,
the pairing gap and the pseudogap show opposite temperature
dependence (Fig.~4c), a typical feature of coexisting but competing
order.
  \singlefig
   {label} % Label
           {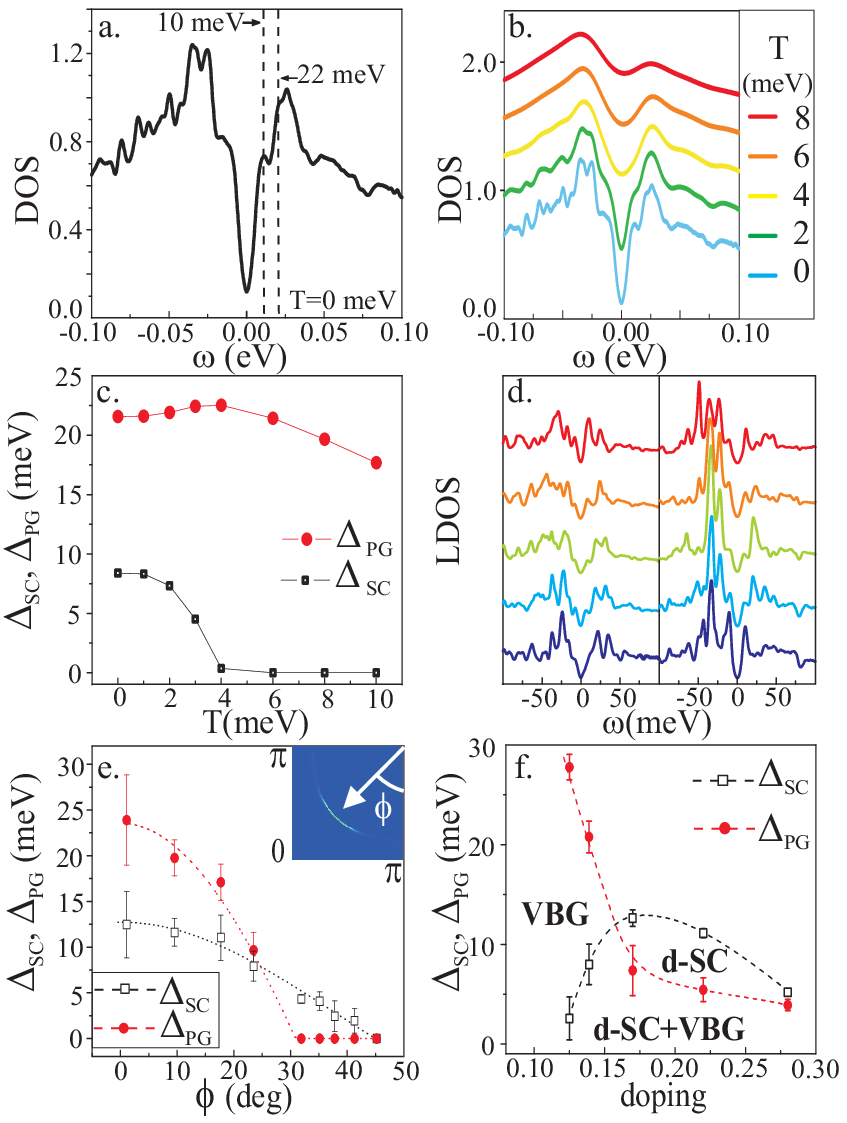} % Filename (with path if needed)
           {0pt} % Extra space above (with units)
           {0.0pt} % Extra space below (with units)
           {1.0} % Scale factor (1.0 is full size)
           {SC state at $x=0.14$. (a) and (b) Averaged DOS showing SC
           gap and VBG pseudogap. (c) Temperature dependence of
           $\Delta_{sc}$ and $\Delta_{pg}$.
           (d) LDOS along two line cuts. (e) Angular
           dependence of SC gap and pseudogap. (f) $\Delta_{sc}$
           and $\Delta_{pg}$ as a function of doping.} % Figure caption

In Figs.~4d, the local DOS (LDOS) is shown along two line cuts on a
typical sample at $x=0.14$ . The evolution of the line-shape agrees
with the STM conductance spectra observed in La-Bi2201
\cite{boyer,vidya}, exhibiting two coexisting low energy gaps that
are spatially inhomogeneous. The momentum dependence of the two
gaps, calculated from the single-particle spectral function
$A(k,\omega)$, is plotted along the underlying FS in Fig.~4e.
Remarkably, the $d$-wave pairing gap extends beyond the Fermi arc
into the antinodal regime. The coexistence of the two gaps off the
Fermi arc in momentum space is an important prediction of the
present theory. Physically, this is a consequence of the softness of
the normal state pseudogap discussed above which allows pairing of
the antinodal states inside the pseudogap. The ground state below
$T_c$ is thus a coherent mixture of $d$-wave VBG and SC pairs; there
are traits of the glassy valence bond order in the pairing gap near
the antinode and {\it vice versa}. Although incoherent background
and inelastic life-time broadening tend to mask the coherent peak
and the pairing gap near the antinode in the spectral function,
recent high resolution ARPES experiments on La-Bi2201 indeed observe
two gaps near the antinodes \cite{vidya,donglaifeng}.

In summary, we presented a VBG theory for the essential features of
the normal state pseudogap and the two-gap phenomena in the SC
state. In Fig.~4f, we construct a theoretical phase diagram using
the doping dependence of the $d$-wave pairing gap and the VBG
pseudogap. It captures the basic topology of the global phase
diagram of the high-T$_c$ cuprates. Although the present theory does
not include a precursor pairing induced pseudogap above $T_c$, it
does not rule out such a possibility due to fluctuations beyond the
Gutzwiller theory.

We thank H. Ding, V. Madhavan, S. Zhou, and C. Li for useful
discussions. This work was supported in part by DOE grant
DE-FG02-99ER45747 and NSF grant DMR-0704545.

\bibliographystyle{unsrt}

\end{document}